\documentstyle[11pt,newpasp,twoside,epsf]{article}
\begin{document}

\title{The Dependence of Tidally-Induced Star Formation on Cluster Density}
\author{C. Moss}
\affil{Vatican Observatory Research Group, Steward Observatory,
University of Arizona, Tucson, AZ 85721, USA}
\author{M. Whittle}
\affil{Department of Astronomy, University of Virginia,
Charlottesville, VA 22903, USA}

\begin{abstract}
A survey of H$\alpha$ emission in 320 spiral galaxies in 8 nearby
clusters shows an enhancement of circumnuclear starburst emission with
increasingly rich clusters.  These observations provide convincing
evidence that spirals have been transformed into S0s in clusters
predominantly by tidal forces, a picture fully in accord with the most
recent numerical simulations of clusters.  For the richest clusters,
the enhancement of starburst emission is greater than would be
expected on the basis of increasing galaxy surface density alone,
which may explain the anomalous result for the type--galaxy surface
density (T--$\Sigma$) relation found for low richness clusters at
intermediate redshift.

\end{abstract}

 
\section{Introduction}
\label{intro}

The remarkable changes in cluster disk galaxy populations between
intermediate redshifts (z $\sim$ 0.5) and the present are well known.
Up to 50\% of the population of intermediate redshift clusters are
comprised of blue, star-forming galaxies, which have been shown to be
predominantly normal spiral and irregular galaxies, a fraction of
which are interacting or obviously disturbed.  By the present epoch,
this population has been depleted by a factor of 2 and replaced by a
population of S0 galaxies.  However the processes by which this has
occurred are still not fully understood (cf. Dressler 1980; Dressler
et al. 1997)

The processes which cause the transformation of the cluster spiral
galaxy population to S0s are expected to have very significant effects
on cluster galaxy star formation rates (SFRs).  We have undertaken
a comparison of SFRs between field and cluster spirals in 8 low
redshift clusters in order to investigate whether these SFRs show
evidence of continuing morphological transformation of disk galaxies
at the present epoch. The comparison survey, details of which are
published elsewhere (Moss, Whittle \& Irwin 1988; Moss \& Whittle
1993; Moss, Whittle \& Pesce 1998; Moss \& Whittle 1999) uses
H$\alpha$ emission, resolved into disk and circumnuclear emission, as
an estimator of the SFRs.

\section{Comparison of Star Formation Rates in Cluster and Field Spirals}

Our cluster sample (viz. galaxies in 8 clusters Abell 262, 347, 400,
426, 569, 779, 1367 and 1656), and our field sample (viz. galaxies in
adjacent supercluster fields) were observed in an identical manner,
thus eliminating systematic effects in detection efficiency.
Furthermore, both (supercluster) field and cluster samples
approximated volume limited samples, largely eliminating the
systematic bias between cluster and field detection rates which may
have been present in many earlier comparison studies (cf. Biviano et
al. 1997).

A difficult question is which criterion to choose to normalise field
and cluster disk galaxy samples. Some authors (e.g.  Hashimoto et
al. 1998; Balogh et al. 1998) have chosen to use bulge to disk (B/D)
ratio on the grounds that it is a less subjective and star formation
contaminated normalisation parameter.  However the relation between
B/D ratio and Hubble T-type has considerable scatter (Baugh, Cole \&
Frenk 1996; Simien \& de Vaucouleurs 1986; de Jong 1995) such that an
increase in the S0/S ratio in clusters is likely to mask systematic
changes of SFR between field and cluster spirals.  Since the latter
changes are of interest for the present study, the B/D ratio is not a
suitable normalisation parameter.

Accordingly we have chosen Hubble type as the normalisation parameter,
and further restricted field and cluster samples to a total of 320
spirals (Sa and later) and peculiars. Some 39\% of spirals and 75\% of
peculiars were detected in H$\alpha$ emission.  It is estimated that
emission detection is 90\% complete to an equivalent width limit of 20
\AA, and $\sim$ 29\% efficient below this limit (cf. Moss et al. 1998).
The detected emission divides approximately equally between {\it
diffuse} and {\it compact} emission which is identified with disk
emission and circumnuclear starburst emission respectively.  Examples
of both types of emission are given in Figure 1.

A particular difficulty which arises in adopting the Hubble type as
the normalisation parameter is the relation between the star formation
properties of the galaxy and its Hubble type.  In
particular, a decrease in disk star formation rate may shift a galaxy
to an earlier type. This shift in type may not be detected in any
comparison of field and cluster spirals (Hashimoto et al. 1998).  This
makes any comparison of {\it disk} emission between field and cluster
spirals uncertain.  By contrast, circumnuclear emission is relatively
independent of type (Kennicutt 1998), and in this case a reliable
comparison of field and cluster spirals is possible.

As is characteristic of circumnuclear starburst emission
(cf. Kennicutt 1998), the detected compact emission correlates with
both a disturbed morphology of the galaxy (significance level,
8.7$\sigma$) indicative of tidally-induced star formation, and with
the presence of a bar (significance level, 3.1$\sigma$).  Furthermore
this emission correlates with both local galaxy surface density
(significance level, 3.9$\sigma$) and cluster central galaxy space
density (significance level, 5.3$\sigma$).  Since there is no
significant difference in the incidence of galaxy bars between the
field and cluster samples, whereas disturbed galaxies are more common
in the cluster environment, it is considered that the observed
enhancement of circumnuclear emission is due to tidally-induced star
formation, whether from galaxy--galaxy, galaxy--group or
galaxy--cluster interactions.

Finally, the enhancement of circumnuclear starburst emission with
increasing cluster density is not wholly accounted for by the
correlation of this emission with local galaxy surface density. A
Kendall partial rank correlation test shows an additional 'cluster
effect' (significance level, 3.3$\sigma$) such that there is a
higher incidence of circumnuclear emission for galaxies in a region
of a given local galaxy surface density in richer clusters, as compared
to that for galaxies in regions of the same surface density in poorer
clusters.  

\begin{figure}[t]
\plottwo{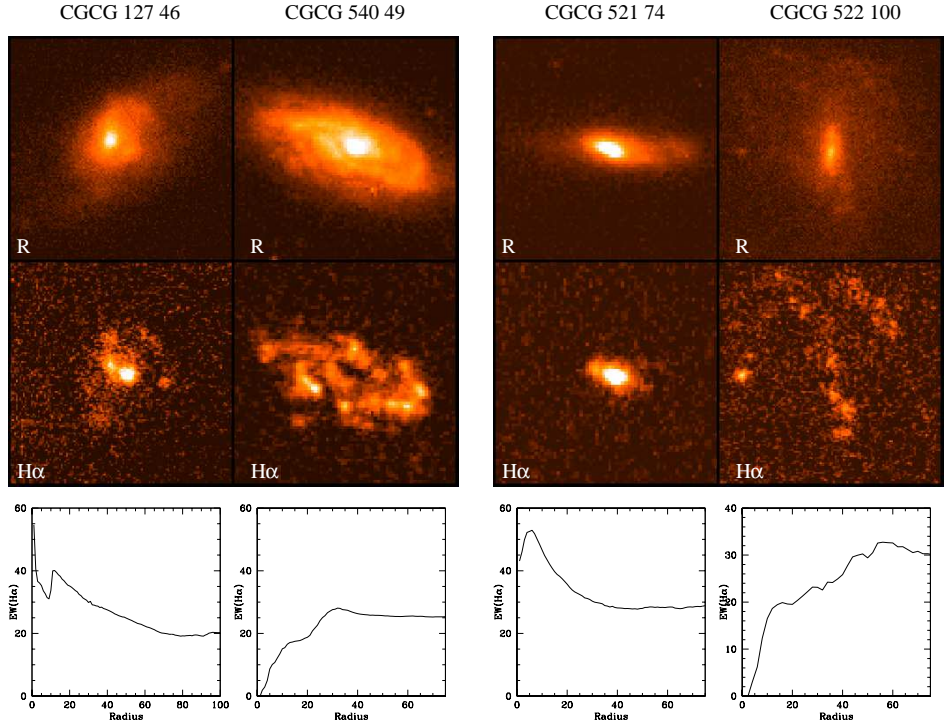}{figure2.epsi}
\caption{R band and H$\alpha$ images for 4 detected galaxies in 
the cluster survey.  The cumulative H$\alpha$ equivalent width with
radial distance from the galaxy center is shown below each pair of
galaxy images. This decreases with radial distance for compact
(circumnuclear starburst) emission (CGCG nos. 127-046, 521-074), whereas it
increases for diffuse (disk) emission (CGCG nos. 540-049, 522-100).}
\end{figure}

\section{Discussion}

Although Lavery \& Henry (1988) first proposed that the Butcher-Oemler
effect could be explained as star formation triggered by
galaxy--galaxy interactions in intermediate redshift clusters, it was
long considered that the typical cluster velocity dispersion ($\sim$
1000 ${\rm km}$ ${\rm s}^{-1}$) was too high for strong tidal interactions
to occur.  However recent work (e.g. Gnedin 1999) has shown that a
non-static cluster potential can enhance tidal interactions for
cluster galaxies.  Such a non-static potential can arise in
sub-cluster merging, and indeed there is evidence that the two richest
clusters in our sample (Abell 1367 and 1656) are recent
post-merger systems (Donnelly et al. 1998; Honda et al. 1996).  It
appears that the merger events in these clusters leading to a rapidly
varying cluster potential, may have caused an increase in galaxy tidal
interactions and the associated observed enhancement of circumnuclear
starburst emission.

Tidal interactions of galaxies in clusters are likely to be an
effective mechanism for the transformation of spirals to S0 galaxies
(e.g. Gnedin 1999).  It was noted above that the observed enhancement
of circumnuclear emission in spirals with increasing cluster density
is not wholly accounted for by that due to increasing local galaxy surface
density.  This implies that tidal interactions, and associated
morphological transformation of spirals to S0s, proceeds faster in
richer as compared to poorer clusters, perhaps because sub-cluster
merging is more common for the former. This in turn may explain the
hitherto anomalous absence of a type--local galaxy surface density
($T-\Sigma$) relation in irregular clusters at intermediate redshift
(Dressler 1980; Dressler et al. 1997). Whereas significant morphological
transformation of the cluster disk galaxy population may be expected
for regular (rich) clusters at $z \sim 0.5$ (for which the timescale
for transformation is shorter) and for irregular (poor) clusters at $z
\sim 0$ (for which a longer time duration for transformation is
available),  for irregular clusters at $z \sim 0.5$, there
has been insufficient time for this
to take place, leading to the observed absence of a
$T-\Sigma$ relation.

\acknowledgements {We thank S.M. Bennett for preparation of the Figure.}

\end{document}